\documentclass[global]{svjour}
\usepackage{graphics}
\journalname{Solar Physics}
\begin{document}

\title{Flux and field line conservation in 3--D nonideal MHD flows}
\subtitle{Remarks about criteria for 3--D reconnection
   without magnetic neutral points}

\author{D.H. Nickeler\inst{1,2} \and H.-J. Fahr\inst{2}
  }

\offprints{D.H. Nickeler}
\mail{D.H.Nickeler@phys.uu.nl}

\institute{Astronomical Institute, University of Utrecht, Princetonplein 5, 3584 CC Utrecht, the
                 Netherlands
           \and
           Institute for Astrophysics and Extraterrestrial Research, Auf dem H\"{u}gel 71, 53121 Bonn,
                 Germany \\
      }

\date{Received: date / Revised version: date}

\maketitle

\begin{abstract}
We make some remarks on reconnection in plasmas and want to present some calculations related
to the problem of finding velocity fields which conserve magnetic flux or at least magnetic field lines. 
Hereby we start from views and definitions of ideal and non-ideal flows on
one hand, and of reconnective and non-reconnective plasma dynamics on the other hand. 
Our considerations give additional insights into the discussion on violations of
the frozen--in field concept which started recently with the papers by Baranov \& Fahr
(2003a;2003b). We find a correlation between the nonidealness which is
given by a generalized form of the Ohm's law and a general transporting velocity, which
is field line conserving.
\end{abstract}

\section{Introduction}\label{intro}

For many applications in astrophysics it is interesting to ask by which velocity the magnetic flux is
transported. This problem of MHD subject was already analysed in several articles, e.g.
Newcomb (1958), Vasyliunas
(1972), Schindler \& Hesse (1988), Hesse \& Schindler (1988),
Hornig \& Schindler (1996). Here we shall make use of the basic ideas of these articles and apply
these to the corresponding more specific problem in heliospheric interface physics.
If no appropriate velocity field for the magnetic flux transport can be found in a plasma region then one can
identify the occurence of magnetic reconnection. The existence of a flux-transporting velocity field is
connected with the \lq ideal\rq\, constraint, that $\vec\nabla\times\vec R=\vec 0$, where $\vec R=\vec E+
\vec {\rm v}\times\vec B$ denotes the nonidealness, $\vec E$, $\vec{\rm v}$ and $\vec B$ being the electric field,
the plasma bulk velocity, and the magnetic field, respectively. The above requirement should be fulfilled 
everywhere in the regarded domain, with the possible exception of some localized regions of non-vanishing
resistivity. Inside the resistive domain the plasma velocity needs 
not to be flux-conserving, but, under certain circumstances, a velocity field different from $\vec {\rm v}$ could
perhaps be found which does the freezing-in of $\vec B$. Such a new velocity field therefore, should be
continuous over the whole domain and converge to the normal flux conserving plasma velocity
outside the resistive domain.

\begin{figure}
\resizebox{0.75\textwidth}{!}{\includegraphics{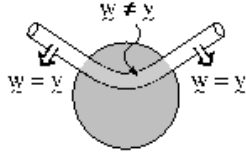}}
\caption{The part of a field line (magnetic flux tube) that is penetrating a nonideal region (shown as the 
   grey shaded sphere) experiences a
   different velocity than the rest of the field line that is still outside.
   The figure is taken from Priest et al. (2003).}
\label{fig:1}
\end{figure}

\begin{figure}
\resizebox{0.75\textwidth}{!}{\includegraphics{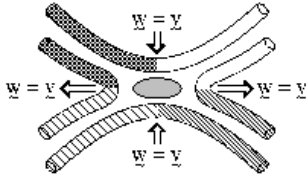}}
\caption{Field lines passing through a nonideal region can disconnect. Outside the nonideal region they
   move with the same smoth velocity field. $\vec w$ is the flux transporting velocity, $\vec {\rm v}$
   is a species bulk flow velocity or plasma velocity and tilde means vector. The figure is taken
   from Priest et al. (2003).}
\label{fig:2}
\label{weqv}
\end{figure}

If it is not possible to find such a new velocity field, $\vec w$ (where $\vec w\equiv$~flux velocity), 
then magnetic reconnection is taking place
since the flux transporting velocity field is not continuous at the border
between the ideal and the nonideal region. This is shown schematically in Fig.\,\ref{fig:1}, where the
flux velocity, $\vec w$, within the nonideal region is different from the plasma velocity, $\vec {\rm v}$.
Outside the nonideal region, the flux velocity, $\vec w$, equals the plasma velocity, $\vec {\rm v}$, as
shown in Fig.\,\ref{fig:2}. Therefore, the nonideal region seems to have disconnected flux tubes.
This means that every magnetic reconnection process is a nonideal process, but not vice versa.

Here we want to answer the question, how the MHD plasma velocity field can be
redefined, in such a way, that it is again the virtue of a flux- or at least a line-conserving flow.
One actual reason for that discussion is the recent dispute between Baranov \& Fahr (2003a;
2003b) and Florinski \& Zank (2003). These authors discussed the meaning of a
generalized Ohm's law, where the nonideal part describes the interaction of different particle species
in a partially ionized plasma.
Baranov and Fahr claimed, that the magnetic field, under specific conditions valid in the heliospheric
interface region, may neither be frozen into the mass-weighted plasma bulk flow,
nor in the ion velocity. We can show very easily, that with the assumptions and the form
of Ohm's law the authors have found (Baranov \& Fahr, 2003a), a flux conserving form
of Ohm's law  could in principle be constructed. However, the needed flow velocity is not necessarily
identical with one of the species bulk velocities, as proposed by Florinski \& Zank (2003),
but critisized by Baranov \& Fahr (2003b).
Florinski \& Zank claim, that the magnetic flux is frozen into the ion velocity within and in the vicinity
of the heliosphere. Baranov and Fahr have shown, that in fact this depends very sensitively on the
solution of the whole set of the multifluid equations, and especially on the magnetic field structure.

We want to show how flux or line conserving velocities could be calculated for a general form of an Ohm's law.
In addition we discuss how the technical procedure is practised and how the included considerations can
principally be used to solve this problem for a partially ionized plasma like that of the
heliospheric interface.

\section{Derivation of flux conserving velocity fields}\label{sec:1}

There are two basic concepts of ideal, respectively non ideal processes in MHD: that of line conservation
(violation), which marks a breakdown of magnetic line connectivity, and that of magnetic flux conservation.
If a highly localized breakdown or violation process takes place in an ideal plasma environment, this
is called magnetic reconnection. However,
even in a nonideal plasma it is possible to get solutions of the nonideal MHD equations, conserving
magnetic flux. For this to happen it is only necessary that 
\begin{equation}
\vec\nabla\times\vec R = \vec 0
   \end{equation}
for any arbitrary closed fluid line (see Priest \& Forbes,\,2000).
The nonidealness $\vec R$ is given by 
 \begin{equation}
\vec R\equiv \vec E + \vec {\rm v}\times\vec B 
   \end{equation}
An explicit discussion of nonideal terms can be found, e.g. in Schl\"uter
(1958).
The second concept of a frozen--in magnetic field is not as strict as the upper one and only requires
\begin{eqnarray} 
 \vec B\times\left( \vec\nabla\times\vec R  \right) = \vec 0
  \end{eqnarray}
which is only fulfilled, if
\begin{equation}
 \vec\nabla\times\vec R = \lambda \vec B 
 \end{equation}
where $\lambda$ is a function of location and time in general and $\vec B\cdot\vec \nabla \lambda = 0$.
If therefore 
\begin{equation}
  \vec B\times\vec\nabla\times\left(\vec E + \vec{\rm v}\times\vec B \right)\neq \vec 0
  \end{equation}
is valid, magnetic field line connection is breaking down and magnetic reconnection is taking
place.

In the stationary case ($\vec\nabla\times\vec E=\vec 0$) the criterion can be written as
\begin{equation}
 \vec B\times \vec\nabla\times\left( \vec{\rm v}\times\vec B  \right)\neq\vec 0  
  \end{equation}
For this reason we want to find a flux conserving velocity field $\vec w$ and a function $X$ with
 \begin{eqnarray}
  \vec E + \vec {\rm v}\times\vec B = \vec R\quad\Leftrightarrow\quad\vec E + \vec w\times\vec B =
   \vec\nabla X\label{genohm}
   \end{eqnarray}
Eq.\,(\ref{genohm}) can be written as
 \begin{eqnarray}
  \left(\vec w - \vec {\rm v}\right)\times\vec B = \vec\nabla X - \vec R
   \end{eqnarray}
Therefore we can formulate
{\theorem: If Ohm's law is given by $\vec E + \vec{\rm v}\times\vec B = \vec R$,
where $\vec{\rm v}$ is the plasma velocity, then the magnetic flux is frozen--in
with respect to velocity fields 

\begin{eqnarray}
 \vec w = \vec{\rm v} +  \frac{\vec B\times\left( \vec\nabla X - \vec R\right)}{|\vec B|^2} + \mu\vec B
  \label{newvelo}\end{eqnarray}
or, equivalently
\begin{eqnarray}
 \vec w = \frac{\left(\vec E - \vec\nabla X\right)\times\vec B}{|\vec B|^2} +\tilde\mu\vec B
  \label{newvelo3}\end{eqnarray}
where $\mu$, i.e. $\tilde{\mu}$ is any arbitrary function in space and $X$ is the solution of the partial
differential equation
\begin{equation}
  \vec B\cdot\vec\nabla X = \vec R\cdot\vec B\label{newvelo2}
  \end{equation}
with appropriate boundary conditions}.
This procedure is reasonable, although it can be seen in the above derivation that even in the
ideal case there exists an infinite number of alternative velocity fields in all of which the magnetic field could 
be frozen in. This additional velocity component is directed in the \lq binormal\rq direction of the
magnetic field, writing $\delta\vec{\rm v}\parallel\vec B\times\vec\nabla X$.
This follows due to $\vec B\cdot\vec\nabla X\equiv 0$, so $X$ is an integral of $\vec B$ (in the ideal case).
That field line motion is not unique, not even in the case of an ideal plasma, was also discussed, e.g. in
Vasyliunas (1972). The bulk velocity field or plasma velocity can therefore be regarded
as a minimal flux preserving velocity (see Vasyliunas, 1972).
Therefore the function $X$  in equation (\ref{newvelo}) should be chosen carefully
(also the term parallel to the magnetic field $\mu\vec B$). With respect to physical
interpretation $\vec w$ should be a velocity field which has a reasonable physical meaning.
Under certain circumstances, the magnetic flux in a partially ionized plasma (see the discusssion
in Baranov \& Fahr\,(2003a; 2003b)) nearly is frozen into the ion velocity.
On the other hand, in a nearly completely ionized plasma electrostatic turbulences could lead to a strong 
localization of the nonidealness (or resistivity in certain cases), so that the flux transporting velocity should 
at least smoothly converge into the normal
known plasma bulk velocity outside or to say far away from the localized nonideal region, which is typical for
astrophysical plasmas. This can be guaranteed by the boundary condition  
\begin{eqnarray}
  X=\textrm{constant}\quad\forall\,\vec x \in D\setminus D_{R}\quad\textrm{and especially on}
  \quad\partial D_{R}\label{bound1}
   \end{eqnarray}
where $D_{R}$ is the nonideal domain. If, however, the magnetic field is nowhere frozen in the bulk fluid motion,
then one has to drop condition (\ref{bound1}), and the plasma is everywhere frozen in the velocity field
(\ref{newvelo}). Then the nonidealness is not localized, and one can hardly speak of a localized nonideal
instability or a localized reconnection process.

Let us now consider a situation in which $\vec R$ vanishes outside a certain domain $D_{R}\subset D$
(or goes faster to zero or to a certain limit). Then we can take equation (\ref{newvelo2}), and use
the identities
\begin{eqnarray}
 \vec E\cdot\vec B &=&\vec R\cdot\vec B \\
  \vec\nabla\cdot\left( X\vec B  \right) &=& \vec B\cdot\vec\nabla X
   \end{eqnarray}
to write down
\begin{eqnarray}
 \int\limits_{D}^{ }\, \vec B\cdot\vec\nabla X  \, dV &=& \int\limits_{D}^{ }\, \vec R\cdot\vec B  \, dV 
  \\
    \nonumber\\
\Rightarrow\quad  \int\limits_{D}^{ }\,\vec\nabla\cdot\left( X \vec B\right)
                      \, dV &=& 
                      \int\limits_{D}^{ }\, \vec E\cdot\vec B  \, dV
\\
\nonumber\\
  \Rightarrow\quad\quad\,\,\,\int\limits_{\partial D}^{ }\, X\,\vec B \cdot d\vec{S} &=&
  \int\limits_{D_{R}}^{ }\, \vec E\cdot\vec B  \, dV \label{hesse}
 \end{eqnarray}
A similar discussion was done by Priest et al. (2003). These authors emphasized the
importance of the component of the electric field $\vec E_{\parallel}$ being aligned with the magnetic field.
Here we take a look to this problem from a different point of view, which is also connected with
$\vec E_{\parallel}$, but emphasizing the role of the Dirichlet boundary condition for $X$.
From this we can see that taking the boundary condition (\ref{bound1}), there is only a
possibility to solve Eq.\,(\ref{newvelo2}), if and only if the right hand side integral over the domain
$D_{R}$ vanishes identically.
This would imply, that the different field aligned parts of the electric fields have to cancel out \lq
statistically\rq\,, e.g. as a result of a strongly spatially fluctuating electric field component due to
turbulence with no preference in direction with respect to the magnetic field. If there is no
preferred direction inside the resistive domain there is at least one surface with vanishing
$\vec E\cdot\vec B=0$, where the Lorentz invariant changes its sign. 
If, in contrast, this integral on the right hand side of Eq.\,(\ref{hesse}) does not vanish and has a non
negligible value, there is no possibility to find a velocity field which exists everywhere in the whole 
domain and which is continuous across the border of the resistive region 
$D_{R}$. Field lines crossing both, the resistive and the
ideal region, are convected with a velocity field, which is not identical with the plasma velocity,
even outside $D_{R}$, and this implies serious breakdown of magnetic flux conservation.
The reason for this is, that field lines being outside and crossing $D_{R}$ have
different flux velocities compared to field lines, which do not cross $D_{R}$ and being completely
outside $D_{R}$.
That is, what can be called reconnection.

 \section{Derivation of line conserving velocity fields}\label{sec:2}

{\theorem: If \,\, $\vec E + \vec{\rm v}\times\vec B =\vec R$\,\, is generalized Ohm's law, then
for ordered, non--ergodic fields, it is always possible
for each solution of the above equation to find a velocity field $\vec w$ and a function $\lambda$
so that 
\begin{eqnarray}
 &&\vec B\times\vec\nabla\times\left(\vec E + \vec w\times\vec B\right) = \vec B\times\vec\nabla\times\vec Y
 = \vec 0 \\\Longleftrightarrow\quad && 
 \vec\nabla\times\vec Y = \lambda\vec B  
    \end{eqnarray} 
and 
\begin{equation}
  \vec w =  \vec{\rm v} +  \frac{\vec B\times\left(\vec Y - \vec R\right)}{|\vec B|^2} + \mu\vec B
\label{newvelo4}
  \end{equation}
where $\vec Y$ has to fullfill the equations
 \begin{eqnarray}
  \vec B\cdot\left(\vec R - \vec Y \right) &=& \vec 0\label{newvelo5}\\
   \vec\nabla\times\vec Y &=& \lambda\vec B\quad\Rightarrow\quad \vec B\cdot\vec\nabla\lambda = 0\, .
    \label{newvelo6}
     \end{eqnarray}

  The existence of fields $\vec Y$, fulfilling Eq.\,(\ref{newvelo6}), together with the existence of a
  corresponding field line constant $\lambda$ imply
  magnetic field line conservation (see e.g. Vasyliunas, 1972). Line conservation means,
  that two fluid elements are always connected by one field line during the convective evolution of
  the electromagnetic field and the velocity field.}

Proof:

We are searching for
\begin{eqnarray}
 &&\vec E + \vec{\rm v}\times\vec B = \vec R\quad\Leftrightarrow\quad\vec E + \vec w\times\vec B = \vec Y\\
 \Rightarrow\quad && \left( \vec w - \vec{\rm v} \right)\times\vec B = \vec Y - \vec R
    \label{newvelo7}
     \end{eqnarray}
%
If one inverts the left hand side of Eq.\,(\ref{newvelo7}) we get
Eq.\,(\ref{newvelo4}) and as necessary additional conditions for $\vec Y$
Eq.\,(\ref{newvelo5}) with line conservation condition Eq.\,(\ref{newvelo6}).

Do these fields $\lambda, \vec Y$ exist? If we use Euler--potentials
\begin{equation}
 \vec Y = Y^{\alpha}\vec\nabla\alpha + Y^{\beta}\vec\nabla\beta  + Y^{s}\vec\nabla s\label{ups}
  \end{equation}    
and take the curl of Eq.\,(\ref{ups}), we get
\begin{eqnarray}
   \vec\nabla\times\vec Y &=&  \left[ \frac{\partial Y^{\beta}}{\partial\alpha} - \frac{\partial Y^{\alpha}}
   {\partial\beta}   \right]
   \left(\vec\nabla\alpha\times\vec\nabla\beta\right) 
     + \left[ \frac{\partial Y^{\alpha}}{\partial s} - \frac{\partial Y^{s}}{\partial\alpha}  
     \right]
     \left(\vec\nabla s\times\vec\nabla\alpha\right)\nonumber\\ 
    && +  \left[ \frac{\partial Y^{s}}{\partial\beta} - \frac{\partial Y^{\beta}}{\partial s}  
   \right]
   \left(\vec\nabla\beta\times\vec\nabla s\right) \nonumber\\
    &&\nonumber\\
    &\stackrel{!}{=}& \lambda(\alpha, \beta)\left( \vec\nabla\alpha\times\vec\nabla\beta \right)\, ,
     \label{fieldcon}
      \end{eqnarray}
 where $\lambda$ in Eq.\,(\ref{fieldcon}) is constant on field lines (see Eq.\,(\ref{newvelo6}))
 and depends therefore on $\alpha$ and $\beta$ only.
 This results in the following set of partial differential equations
 \begin{eqnarray}
   \qquad &&   \frac{\partial Y^{\beta}}{\partial\alpha} - \frac{\partial Y^{\alpha}}{\partial\beta} =
                     \lambda(\alpha, \beta)
                     \label{pde1} \\ \nonumber\\
                &&    \frac{\partial Y^{\alpha}}{\partial s} - \frac{\partial Y^{s}}{\partial\alpha} = 0
                      \label{pde2}\\ \nonumber\\
                &&     \frac{\partial Y^{s}}{\partial\beta} - \frac{\partial Y^{\beta}}{\partial s} = 0
                       \, . \label{pde3}
    \end{eqnarray}
Equation (\ref{newvelo5}) and $\vec B\cdot\vec\nabla s \equiv \left( \vec\nabla\alpha\times\vec\nabla\beta
\right)\cdot\vec\nabla s = |\vec B|$ lead to 
\begin{eqnarray}
  0 = && \vec B\cdot\left( \vec R - \vec Y \right)\\
  = && \left(\vec\nabla\alpha
     \times\vec\nabla\beta\right)
     \cdot\left[ (R^{\alpha} - Y^{\alpha})\vec\nabla\alpha + (R^{\beta} - Y^{\beta})
     \vec\nabla\beta \right.\nonumber\\
      && \bigl. + (R^{s} - Y^{s})\vec\nabla s \bigr]\\
  = &&\vec B\cdot\vec\nabla s \left( R^{s} - Y^{s} \right) = |\vec B|\left( R^{s} - Y^{s} \right) \\
   \Rightarrow && R^{s}= Y^{s}\label{electrics}
        \end{eqnarray}
due to $\vec R = R^{\alpha}\vec\nabla\alpha + R^{\beta}\vec\nabla\beta + R^{s}\vec\nabla s$.
The integration procedure is similar to that done in Hesse \& Schindler (1988).
The solution of the differential Eqs.\,(\ref{pde2}) and (\ref{pde3}) can be formally
expressed by
 \begin{eqnarray}
  Y^{\alpha} &=& \int  \frac{\partial R^{s}}{\partial\alpha} \, ds + \chi^{\alpha}(\alpha,\beta)
  \label{construct}\\
   Y^{\beta} &=& \int  \frac{\partial R^{s}}{\partial\beta}\, ds + \chi^{\beta}(\alpha,\beta)\, ,
    \label{construct1}
     \end{eqnarray} 
so that we get a \lq parametric\rq~dependence of the vector $\vec Y$ upon the nonidealness
$\vec R$ and therefore a dependence of the new velocity field $\vec w$ upon the nonidealness
$\vec R$, in contrast to the formulation in Hesse \& Schindler (1988).
Setting the ansatz Eqs.\,(\ref{construct}) and (\ref{construct1}) into Eq.\,(\ref{pde1}), we get
 \begin{eqnarray}
   \frac{\partial\chi^{\beta}}{\partial\alpha} - \frac{\partial\chi^{\alpha}}{\partial\beta}
    \stackrel{!}{=}\lambda\, .
     \end{eqnarray}
 The formal representation of $\chi^{\alpha}$ and $\chi^{\beta}$ reads
      \begin{eqnarray}
        \chi^{\alpha}:=\int \lambda^{\alpha}\, d\beta\quad\textrm{and}\quad
         \chi^{\beta}:=\int \left(\lambda + \lambda^{\alpha}\right)\, d\alpha\, ,
          \end{eqnarray}
 where $\lambda^{\alpha}$ is a function of $\alpha$ and $\beta$ only.
 From Eqs.(\ref{construct}) and (\ref{construct1}) it can be seen, that $Y^{\alpha}$ and $Y^{\beta}$
 cannot vanish along a field line, as the second terms in both equations depend on $\alpha$ and
 $\beta$ only. If $\vec R$ is strictly localized, both components cannot vanish in the
 direction along the field line, passing through the nonideal region. 
 These components are constant on field lines, and therefore there will be an additional velocity
 component to the flux velocity, $\vec B\times\vec Y$, which is perpendicular to the magnetic field. But if
 these field lines are not passing through the nonideal region, one can see, that $\lambda$ vanishes outside
 the flux tube.
 In this case, the field $\vec Y$ could be written as a gradient. This enables us to find
 a common field line velocity for this field lines. 

 For all field lines passing through the localized (around $\alpha_{0},\beta_{0},s_{0}$)
 nonideal region and extending to infinity, the additional velocity component $\vec B\times\vec Y$ will
 not vanish, so that the flux conserving velocity field is discontinuous.

\section{Ohm's law in a partially ionized plasma -- application to the heliosphere}\label{sec:3}

 In front of the heliosphere there is a wall of neutral gas, called the hydrogen wall. This leads to the
 demand, that for describing plasma dynamics in this region, it is important not to use ideal
 Ohm's law, but a more complicated form of Ohm's law, found by Cowling (1976) and
 Kulikovskii (1962). 
 This was also used by Baranov \& Fahr  (2003a;2003b), who discussed its
 importance for heliospheric plasma dynamics.

 We will now show, that with Eq.\,(\ref{hesse}) it is possible for every point in space and at any time
 to find a flux conserving velocity field
 because $\vec E\cdot\vec B = 0 $ everywhere.

 With the following shape of the nonidealness $\vec R$
  \begin{eqnarray}
  \vec R\equiv \frac{(1-\alpha)^2 c^2}{K_{ia}}\left[\frac{\alpha}{1+\alpha}\,\vec\nabla P\times\vec B
  + \frac{1}{4\pi}\,\vec B\times\left(\left(\vec\nabla\times\vec B  \right)\times\vec B\right)  \right]
       \end{eqnarray}
 Ohm's law can be written as
 \begin{eqnarray}
  \vec E + \vec{\rm v}\times\vec B = \vec R\label{ohmpart}
       \end{eqnarray}
Then Eq.\,(\ref{ohmpart}) can be rewritten as
 \begin{eqnarray}
  \vec E + \left[ \vec {\rm v} -  \frac{(1-\alpha)^2 c^2}{K_{ia}}\left[\frac{\alpha\, \vec\nabla P}{1+\alpha}
  - \,\left(\left(\vec\nabla\times\vec B  \right)\times\vec B\right)  \right]\right]
  \times\vec B = \vec 0\, ,
      \end{eqnarray}
where the term in brackets in front of the cross product with $\vec B$ is the flux transporting velocity.
But we see, that the velocity depends explicitly on the structure of the (electro--)magnetic field,
so it is not clear in advance, which should be the precise velocity, and it turns out,
that it is not necessarily a weighted or certain species velocity. This makes it difficult to talk
about a typical velocity field, in which the magnetic flux is frozen in
(see e.g. discussion in Hornig \& Schindler,\,1996). So no distinct species velocity can be
determined as flux conserving velocity.
The detection of magnetic reconnection in such a plasma is only possible if the physical parameters
allow to determine a general ideality: there must exist a flux
transporting velocity almost everywhere.

\section{Discussion and conclusions}\label{sec4}
  
It may be useful finding velocity fields in nonideal environments, in which at least the
  magnetic field lines are frozen in, but maybe not the magnetic flux, so that in some regions magnetic
  flux could be annihilated, diffuse or may be even created by special flows (dynamo). 
  The aim is to identify velocity fields in a plasma/medium, which includes different ions, neutrals
  and maybe dust, for future work. At this point of the discussion we thus can only conclude with the 
  statement that the freezing--in velocity field is a very complicated nonlocally determined field depending
  on many nonlocal properties of the field configuration and the differential neutral gas flow relative to the
  plasma flow.

\end{document}